\documentclass[showpacs,prl,preprintnumbers,amsmath,amssymb]{revtex4}

\usepackage{graphicx}
\usepackage{dcolumn}
\usepackage{bm}

\begin{document}

\title{Inhomogeneous distribution of particles and temperature in a self-gravitating system}

\author{B. I. Lev}

\affiliation{Bogolyubov Institute for Theoretical Physics, NAS of
Ukraine, Metrolohichna 14-b, Kyiv 07680, Ukraine}

\date{\today}

\begin{abstract}
Self-gravitating systems are non-equilibrium \textsl{a priory}. A new approach is proposed, 
which employs a non-equilibrium statistical operator taking into account inhomogeneous 
distributions of particles and temperature. The method involves the saddle-point 
procedure to find the dominant contributions to the partition function and thus to 
obtain all thermodynamic parameters of the system. Probable peculiar features in the 
behavior of the self-gravitating system are considered for various conditions. The equation
of state for self-gravitating system has been determined. A new length of the statistical 
instability and parameters of the spatially inhomogeneous distribution of particles and 
temperature are obtained for a real gravitational system.
\end{abstract}

\maketitle

\section{Introduction}

The study of self-gravitating systems is of fundamental physical importance. 
Such system are suitable for testing the ideas on the statistical description 
of systems with by long-range interaction. Moreover, a self-gravitating system 
concerns a general problem that has been studied for a long time \cite{Pad} 
and turned out to be  more complicated than other many-body system. Thus, 
the self-gravitating system provides a model of fundamental interest for which 
ideas of statistical mechanics and thermodynamics can be tested and developed 
\cite{Chav}. The statistical description of the system can be helpful for the 
problems of astrophysics \cite{Sas},\cite{Chan} and for the development of general  
approaches to the investigation of structure formation in various physical situations.

The statistical mechanics of the systems under consideration turns out to be 
very different from the study of other, more familiar, many-body systems.  
The fundamental difference is that the  concept of equilibrium is not always 
well defined and these systems  manifest nontrivial behavior with  under  phase 
transitions associated with the gravitational collapse. The standard methods of 
the equilibrium statistical mechanic cannot be  applied to the study of self-gravitating 
systems since in this case the thermodynamic ensembles are not equivalent, i.e., 
the negative specific heat \cite{Tir} occurring in the micro-canonical ensemble 
does not exist in the canonical description \cite{Cav}. In the micro-canonical ensemble, 
the self-collapse corresponds to the "gravy-thermal catastrophe"  while in the 
canonical ensemble it is associated with the "isothermal collapse" \cite{Cha}.  
Energy of an isolated self-gravitating system is conserved and fundamentally, only 
the concept of micro-canonical ensemble makes natural sense. In particular, inasmuch 
as energy is not-additive, the canonical ensemble cannot be applied to the study  of 
systems with long-range interactions. 

Two  approaches (statistical and thermodynamical) have been developed to  
find the equilibrium states of the self-gravitating system and to describe  
probable phase transitions \cite{Cav}, \cite{Cha}. The collapse in such systems  
starts as a spatially inhomogeneous distribution of particles in all the  system 
at once. 

A very important question  is whether the inhomogeneous distribution of particles 
can be regarded as an equilibrium state. Phase transitions in such systems  
require the description in terms of the mean-field thermodynamics  \cite{Cha}. 
It is generally believed that the mean field theory is exact for equilibrium  
systems. Since in the mean field theory any thermodynamic function depends  
only on dimensionless combinations of thermodynamic variables, it follows 
from the article \cite{San} and the book \cite{Rue} that the thermodynamic 
limit of the system does not exist. The non-existence of the thermodynamic 
limit implies that the thermodynamic potentials do not scale properly with 
the number of particles and thus the thermodynamic functions, i.e., temperature, 
diverge. Nevertheless,  we can take the ordinary thermodynamic limit and then  
employ the usual thermodynamic tools by first regularizing the long-distance 
behavior of the gravitational potential and then introducing a very large 
screening length. The system is then thermodynamically stable and the thermodynamic 
limit does exist \cite{Lal}.

Formation of  spatially inhomogeneous distributions of interacting particles is a 
typical problem in condensed matter physics and requires non-conventional  
statistical description  in the case of particles involved in the gravitational 
interaction with regard for an arbitrary spatially inhomogeneous particle distribution. 
The statistical description  should employ the procedure to find the dominant 
contributions to the partition function and to avoid entropy divergences for 
infinite system volume. 

A non-conventional method was proposal in \cite{Bel}, \cite{Lev}, \cite{Kle}. 
It employs the Hubbard-Stratonovich representation of the statistical sum \cite{Str} 
that is  extended to a system with gravitational interaction in order to find a 
solution for the particle distribution making no use of spatial box restrictions. 
It is important that this solution has no divergences in the thermodynamic limits. 
For this  purpose the saddle point approximation is applied, which takes into 
account the conservation of the number of particles in the limiting space and  
yields a nonlinear equation. The three-dimensional solution of this nonlinear 
equation selects the states whose contributions in the partition function are 
dominant. The partition function  for the case of homogeneous particle distribution,  
as well as for the case of inhomogeneous distribution, was obtained in \cite{Lev}, 
\cite{Gri}. This approach, however, describes only the condition for the formation 
of probable inhomogeneous distributions of  gravitating particles. 

Systems with long-range interactions, such as self-gravitating system, do not 
relax to the usual Boltzmann-Gibbs thermodynamic equilibrium, but  get trapped 
in the quasi-stationary states whose lifetimes  diverge as the number of particles 
increases. A theory was proposed  that provides a quantitative prediction of the 
instability threshold for the spontaneous symmetry breaking for a class of d-dimensional 
systems\cite{Pak}. The non-equilibrium stationary states of such systems were described 
in  the paper \cite{Ben} where the three-dimensional systems were shown to be trapped in 
non-equilibrium quasi-stationary states  rather than evolve to the thermodynamic equilibrium  
\cite{Ben}, mainly because the self-gravitating systems exist in a highly nonequilibrium 
states and the time of relaxation to the equilibrium state is very long. The homogeneous 
particle distribution in a self-gravitating system is not stable. Particle distributions 
in such  systems  are spatially inhomogeneous, from the very beginning. Therefore, the 
system brakes on a complex of inhomogeneous clusters, which  collapse to more condensed 
states. The description of typical behavior of  a self-gravitating system should be specific  
for various equilibrium ensembles. 

Some attempts to include particle distribution inhomogeneity into consideration have 
been made \cite{Mich}-\cite{Jaf},  however, the solution has not been found up to now. 
The reason is that, with the inhomogeneity included, the chemical potential depends 
on the space variables and the relevant equations of state should interconnect temperature and 
density. Hence the temperature, as a thermodynamic parameter, has to depend on the 
space coordinates too. The concentration-dependence of temperature and is found in 
order to obtain stable solutions for the gravitational formation of stars \cite{Cha}. 
This approach  seems to be  inconsistent, since the state equation should follow from 
the definition of the partition function which, however, is unknown for spatially 
inhomogeneous systems \cite{Bax}, \cite{Rue}. Therefore, there is a dilemma, whether 
to employ the postulates of equilibrium statistical mechanics  and obtain only  
instability criteria or not to try to take into account the  spatial inhomogeneity  
and use a different approach. Such inhomogeneous distributions of  particles, temperature, 
and chemical potential can be accounted for in the non-equilibrium statistical operator 
approach \cite{Zub} with allowance for probable local changes of the thermodynamic 
parameters. This system is non-equilibrium and thus inhomogeneous particle distribution  
can justify the inhomogeneous distribution of temperature, chemical potential, and other
thermodynamic parameters.

In this article we propose a new approach in terms of the non-equilibrium statistical 
operator \cite{Zub} that is are more suitable for the description of  gravitational 
systems. The equation of state and all thermodynamic characteristics needed are defined 
by  equations which govern the greatest contributions to the partition function. Thus, 
there is no need to introduce additional hypothesis about the density-dependence of 
temperature . This dependence  is obtained by solving corresponding thermodynamic 
relations which describe the extremums of the non-equilibrium partition function. 
The possible  spatially inhomogeneous distributions of particles and temperature  
are obtained for simple cases. For the equilibrium case, the well-known result 
\cite{Hua},\cite{Isi}  for the partition function  is reproduced. This approach 
is shown to describe the inhomogeneous particle distribution and to determine the 
thermodynamic parameters in the self-gravitating system. The main idea of this paper 
is to provide a  detailed description of self-gravitating systems  in terms of the 
principles of non-equilibrium statistical mechanics and to obtain  distributions of 
particles and temperature  for fixed number of particles and energy of the system.

\section{Non-equilibrium statistical sum}

Phenomenological thermodynamics is based on the conservation laws for the average 
values of physical parameters, i.e.,  the number of particles, energy, and  momentum. 
Statistical thermodynamics of non-equilibrium systems is also based  on the conservation  
laws, however, for the dynamic variables rather than their  average values. It presents 
local conservation  laws for the dynamic variables. In order to find thermodynamic 
functions of a non-equilibrium system, we have to use the presentation of  relevant 
statistical ensembles taking into account the non-equilibrium states of these systems. 
The conception of Gibbs ensembles can  provide a description of non-equilibrium stationary 
states of the system. In this case we can  define a non-equilibrium ensemble as a 
totality of systems  that can be contained in the same stationary external condition. 
This system  possesses the character of contact  similar to a thermostat and possesses 
all  the probable values of macroscopic parameters  compatible with the given conditions. 
Local equilibrium stationary distributions are formed in systems under similar 
stationary external conditions. If the external condition  depends on time, the  
relevant local equilibrium distribution  is not stationary.  In order to determine 
a local equilibrium ensemble exactly,  we have to determine the distribution function 
or the statistical operator of the system \cite{Zub}. Finally,  we remind the reader 
that stable states of classical self-gravitating particles are only metastable because 
they correspond to the local maxima of the thermodynamic potential. This thermodynamic 
potential  is the local entropy whose extremum determines the behavior of the system.

Under the assumption that non-equilibrium states of the system can be written in terms 
of the inhomogeneous distribution energy, $H(\mathbf{r})$, and the number of particles 
(density), $n\left(\mathbf{r}\right)$, the local equilibrium distribution function for 
a classical system can write in the form \cite{Zub}
\begin{equation}
f_{l}=Q^{-1}_{l}\exp\left\{-\int
\left(\beta(\mathbf{r})H(\mathbf{r})-\eta(\mathbf{r})n(\mathbf{r})\right)d\mathbf{r}\right\}
\end{equation}
where 
\begin{equation}
Q_{l}=\int D\Gamma \exp\left\{-\int
\left(\beta(\mathbf{r})H(\mathbf{r})-\eta(\mathbf{r})n(\mathbf{r})\right)d\mathbf{r}\right\}
\end{equation} 
determine the statistical operator of the local equilibrium distribution. The integration 
in  this formula should be performed over  the whole phase space of the system. It should 
be  noted that in the case of local equilibrium distribution, the Lagrange multipliers 
$\beta(\mathbf{r})$ and $\eta(\mathbf{r})$ are functions of a spatial point. The density 
of particles can be presented in the standard form, i,e,
\begin{equation}
n(\mathbf{r})=\sum_{i}\delta(\mathbf{r}-\mathbf{r_{i}})
\end{equation} 
The local equilibrium distribution  can be introduced provided the relaxation time in 
the whole system is  greater than the relaxation time in a local macroscopic region 
contained in this system.

Having determined the non-equilibrium statistical operator we can obtain all the 
thermodynamic parameters of the non-equilibrium system. To do this we have to derive 
a thermodynamic relation for inhomogeneous systems. The variation of the statistical 
operator by the Lagrange multipliers  yields the required thermodynamic relation in 
the form \cite{Zub}:
\begin{equation}
-\frac{\delta \ln Q_{l}}{\delta \beta(\mathbf{r})}=\langle
H(\mathbf{r})\rangle_{l}
\end{equation}
and
\begin{equation}
\frac{\delta \ln Q_{l}}{\delta \eta (\mathbf{r})}=\left\langle
n(\mathbf{r})\right\rangle_{l}
\end{equation}
This relation is a natural general  extension of the well-known relation  
for equilibrium systems to the case of inhomogeneous system. Conservation 
of the number of particles and energy in the system  is given by
\begin{equation}
\int n(\mathbf{r})d\mathbf{r}=N
\end{equation}
and
\begin{equation}
\int H(\mathbf{r})d \mathbf{r}=E
\end{equation}

To continue the statistical description of non-equilibrium systems,  we have 
to find the Hamiltonian of the system. In the general case, the Hamiltonian of 
a system of interacting particles is given by 
\begin{equation}
H=\sum_{i}\frac{p_{i}^{2}}{2m}+\frac{1}{2}\sum_{i,j}
W(\mathbf{r_{i}}\mathbf{r_{j}})
\end{equation}
The energy of gravitational interaction energy can be written  in the well-known form
\begin{equation}
W(\mathbf{r},\mathbf{r'})=\frac{G m^{2}}{|\mathbf{r}-\mathbf{r'}|}
\end{equation}
where $G$ is the gravitational constant and $m$ is the mass of a separate particle.  
In what follows we use only the density of energy which for a self-gravitating system 
can be written in the form
\begin{equation}
H(\mathbf{r})=\frac{p^{2}(\mathbf{r})}{2m}n(\mathbf{r})+\frac{1}{2}\int
W(\mathbf{r},\mathbf{r'})n(\mathbf{r})n(\mathbf{r'})d \mathbf{r'}
\end{equation}
This energy density  of a system  can be used if we divide the space into fragments 
with equal masses and consider motion in the phase space of an incompressible gravitational 
fluid.  This model is valid for  collision-less systems and particles with gravitational 
interaction i provide a very good example of such system.

The non-equilibrium statistical operator of a self-gravitating system can be written as
\begin{equation}
Q_{l}=\int D\Gamma \exp\left\{-\int
\left(\beta(\mathbf{r})\frac{p^{2}(\mathbf{r})}{2m}-\eta(\mathbf{r})\right)n(\mathbf{r})d\mathbf{r}-\frac{1}{2}\int
W(\mathbf{r},\mathbf{r'})n(\mathbf{r})n(\mathbf{r'})d\mathbf{r}d
\mathbf{r'}\right\}
\end{equation}
Integration over the phase space  is given by 
$D\Gamma=\frac{1}{\left( 2\pi \hbar\right)^{3}}\prod\limits_{i}dr_{i} dp_{i}$. 

In order to perform formal integration in the second part of this  paper, 
we introduce additional field variables  in terms of the theory of Gaussian 
integrals \cite{Str}, \cite{Kle}, i.e.,
\begin{equation}
\exp \left\{ -\frac{1}{2}\int
\beta(\mathbf{r})W(\mathbf{r},\mathbf{r'})n(\mathbf{r})n(\mathbf{r'})d
\mathbf{r} d \mathbf{r'}\right\} = \int D \varphi \exp \left\{
-\frac{1}{2}\int
W^{-1}(\mathbf{r},\mathbf{r'})\varphi(\mathbf{r})\varphi(\mathbf{r'})d
\mathbf{r}d \mathbf{r'}-
\int\sqrt{\beta(\mathbf{r})}\varphi(\mathbf{r})n(\mathbf{r})d
\mathbf{r} \right\}
\end{equation}
where $D\varphi=\frac{\prod\limits_{s}d\varphi _{s} }{\sqrt{\det
2\pi \beta W(\mathbf{r},\mathbf{r'})}}$ and
$W^{-1}(\mathbf{r},\mathbf{r'})$ is the inverse operator  that satisfies the condition
$W^{-1}(\mathbf{r},\mathbf{r'})W(\mathbf{r'},\mathbf{r''})=\delta
(\mathbf{r}-\mathbf{r''})$.  Now the field variable $\varphi(\mathbf{r}) $ contains the 
same information as the original distribution function, i.e.,  complete information about 
probable spatial states of the system. 
The inverse operator $W^{-1}(\mathbf{r},\mathbf{r'})$ of the gravitational interaction 
in the continuum limit should be treated in the operator sense, i.e.,
\begin{equation}
W^{-1}(\mathbf{r},\mathbf{r'})=-\frac{1}{4\pi G
m^{2}}\triangle_{\mathbf{r}}\delta(\mathbf{r}-\mathbf{r'})
\end{equation}
where $\triangle_{\mathbf{r}}$- is the Laplace operator in the real space. The statistical 
operator reduces now to the form
\begin{equation}
Q_{l}=\int D\Gamma \int D\varphi \exp\left\{-\int
\left(\beta(\mathbf{r})\frac{p^{2}(\mathbf{r})}{2m}-\eta(\mathbf{r})-\sqrt{\beta(\mathbf{r}))}\varphi(\mathbf{r}\right)n(\mathbf{r})d\mathbf{r}-
\frac{1}{8\pi m^{2}G}\int \left(\nabla \varphi(\mathbf{r})
\right)^{2}d\mathbf{r}\right\}
\end{equation}
This functional integral can be  integrated over the phase space.  Making use of the 
definition of density we can rewrite the non-equilibrium statistical operator as
\begin{equation}
Q_{l}=\int D\varphi \int \frac{1}{\left( 2\pi \hbar\right)
^{3}N!}\prod\limits_{i}dr_{i} dp_{i}\xi(\mathbf{r_{i}})\exp\left\{-
\left(\beta(\mathbf{r_{i}})\frac{p_{i}^{2}}{2m}-\sqrt{\beta(\mathbf{r_{i}}))}\varphi(\mathbf{r_{i}}\right)-
\frac{1}{8\pi m^{2}G}\int \left(\nabla \varphi(\mathbf{r})
\right)^{2}d\mathbf{r}\right\}
\end{equation}
where $\xi(\mathbf{r})\equiv \exp \eta(\mathbf{r})$ is a new variable that  can be 
interpreted as chemical activity. Now we can  perform integration over  the momentum. 
The non-equilibrium statistical operator  is given by
\begin{equation}
Q_{l}=\int D\varphi \exp\left\{-
\frac{1}{8\pi m^{2}G}\int \left(\nabla \varphi(\mathbf{r})
\right)^{2}d\mathbf{r}\right\}\frac{1}{N!}\prod\limits_{i}\int dr_{i}\xi(\mathbf{r_{i}})\left(\frac{2\pi m}{\hbar^{3}\beta(\mathbf{r_{i}})}
\right)^{\frac{3}{2}}\exp
\left(\sqrt{\beta(\mathbf{r_{i}}))}\varphi(\mathbf{r_{i}}\right)
\end{equation}

\begin{equation}
Q_{l}=\int D\varphi \exp\left\{-
\frac{1}{8\pi m^{2}G}\int \left(\nabla \varphi(\mathbf{r})
\right)^{2}d\mathbf{r}\right\}\sum_{N}\frac{1}{N!}{\int dr \xi(\mathbf{r})\left(\frac{2\pi m}{\hbar^{3}\beta(\mathbf{r})}
\right)^{\frac{3}{2}}\exp
\left(\sqrt{\beta(\mathbf{r}))}\varphi(\mathbf{r}\right)}^{N}
\end{equation}

Now it reduces to the simple form
\begin{equation}
Q_{l}=\int D\varphi \exp\left\{\int \left[-\frac{1}{8\pi m^{2}G}
\left(\nabla \varphi(\mathbf{r})\right)^{2}+
\xi(\mathbf{r})\left(\frac{2\pi m}{\hbar^{3}\beta(\mathbf{r})}
\right)^{\frac{3}{2}}\exp\sqrt{\beta(r)}\varphi(\mathbf{r})\right]d\mathbf{r}\right\}
\end{equation}
For constant temperature $\beta$ and absolute chemical activity $\xi$, the statistical 
operator fully  reproduces the equilibrium canonical partition function \cite{Lev},
\cite{Veg}.In our general case, the non-equilibrium statistical operator can be rewritten 
in the form
\begin{equation}
Q_{l}=\int D\varphi
\exp\left\{-S(\varphi(\mathbf{r}),\xi(\mathbf{r}),\beta(r))\right\}
\end{equation}
where the effective non-equilibrium "local entropy"  is given by
\begin{equation}
S(\varphi(\mathbf{r}),\xi(\mathbf{r}),\beta(r))=\int
\left[\frac{1}{8\pi m^{2}G} \left(\nabla
\varphi(\mathbf{r})\right)^{2}- \xi(\mathbf{r})\left(\frac{2\pi
m}{\hbar^{2}\beta(\mathbf{r})}
\right)^{\frac{3}{2}}\exp\sqrt{\beta(r)}\varphi(\mathbf{r})\right]d\mathbf{r}
\end{equation}
The statistical operator  makes it possible to use the  efficient methods developed 
in the quantum field theory without  any additional restrictions for the integration 
over the field variables or the perturbation theory. The functional $S(\varphi(\mathbf{r}),
\xi(\mathbf{r}),\beta(r))$ depends on the distribution of the field variables 
$\varphi(\mathbf{r}) $, the chemical activity $\xi(\mathbf{r})$, and the inverse 
temperature $\beta(\mathbf{r})$. Now the saddle-point method can  be employed to 
find the asymptotic value of the statistical operator $Q_{l}$ for $N$ to $\infty $; 
the dominant contribution is given by the states which satisfy the extremum conditions 
for the functional.  It is not difficult to see that the saddle-point equation represents 
the thermodynamic relation and  may be reduced to an equation for the field variable, i.e., 
\begin{equation}
\frac{\delta S}{\delta \varphi(\mathbf{r})}=0
\end{equation}
with the normalization condition being given by
\begin{equation}
\frac{\delta S}{\delta (\eta(\mathbf{r}))}=-\int \frac{\delta
S}{\delta (\xi(\mathbf{r}))}\xi(\mathbf{r}))d\mathbf{r}=N
\end{equation},
and for the energy conservation  in the system, i.e.,
\begin{equation}
-\int \frac{\delta S}{\delta
(\beta(\mathbf{r}))}\xi(\mathbf{r}))d\mathbf{r}=E
\end{equation}
The solution of this equation  completely determines all the thermodynamic parameters 
and describes the general behavior of a self-gravitating system both for spatially 
homogeneous and inhomogeneous particle distributions. The above set of equations in 
principle solves the many-particle problem in the thermodynamic limit. The spatially 
inhomogeneous solution of  these equations corresponds to the distribution of interacting 
particles. Such inhomogeneous behavior is associated with the nature and intensity of 
the interaction. In other words, accumulation of particles in a finite spatial region 
(cluster formation ) reflects the spatial distribution of the field,  activity, and 
temperature. It is very important to note that this approach is the only one that makes 
it possible to take into account the inhomogeneity of temperature distribution  that may 
depend on the spatial distribution of particle in the system. In other approaches, the 
dependence of temperature  on a spatial point  is introduced through the polytrophic 
dependence of temperature on particle density in the equation of state \cite{Cha}. 
In the present approach, this dependence follows from the necessary thermodynamic 
condition and can be  found for various particle distributions. 

Now we derive the saddle-point equation  for the extremum of the local entropy 
functional $S(\varphi,\xi,\beta)$. The equation for the field variable 
$\frac{\delta S}{\delta \varphi}=0$ yields
\begin{equation}
\frac{1}{r_{m}}\triangle
\varphi(\mathbf{r})+\xi(\mathbf{r})\left(\frac{2\pi
m}{\hbar^{2}\beta(\mathbf{r})}
\right)^{\frac{3}{2}}\sqrt{\beta(r)}\exp(\sqrt{\beta(r)}\varphi(\mathbf{r}))=0
\end{equation}
where the notation $r_{m}\equiv 4\pi G m^{2}$ is introduced. The normalization condition 
may be written as 
\begin{equation}
\int \xi(\mathbf{r})\left(\frac{2 m}{\hbar^{2}\beta(\mathbf{r})}
\right)^{\frac{3}{2}}\exp(\sqrt{\beta(r)}\varphi(\mathbf{r}))d\mathbf{r}=N
\end{equation}
and the equation for the energy conservation  in the system is given by
\begin{equation}
\frac{3}{2}\int \left(\frac{2\pi m}{\hbar^{2}\beta(\mathbf{r})}
\right)^{\frac{3}{2}}\frac{\xi(\mathbf{r})}{\beta(\mathbf{r})}(3-
\sqrt{\beta(r)}\varphi(\mathbf{r}))\exp(\sqrt{\beta(r)}\varphi(\mathbf{r}))d\mathbf{r}=E
\end{equation}
To draw more information on the behavior of a self-gravitating system,  we introduce 
new variables. The normalization condition $\int \rho(\mathbf{r})d\mathbf{r}=N$  yields 
the definition for  the density function, i.e., 
\begin{equation}
\rho(\mathbf{r})\equiv \left(\frac{2\pi
m}{\hbar^{2}\beta(\mathbf{r})}
\right)^{\frac{3}{2}}\xi(\mathbf{r})\exp(\sqrt{\beta(\mathbf{r})}\varphi(\mathbf{r}))
\end{equation}
which reduces the equation to a simpler form. The equation for the field variable is given by
\begin{equation}
\triangle
\varphi(\mathbf{r})+r_{m}\sqrt{\beta(\mathbf{r})}\rho(\mathbf{r})=0
\end{equation}
The equation for energy conservation takes the form
\begin{equation}
\frac{1}{2}\int \frac{\rho(\mathbf{r})}{\beta(\mathbf{r})}(3-
\sqrt{\beta(r)}\varphi(\mathbf{r}))d\mathbf{r}=E
\end{equation}
The equation thus obtained cannot be solved in the general case, but it is possible to 
analyze many cases  of the behavior of a self-gravitating system  under various external 
conditions. In what follows we write the chemical activity in terms of the chemical potential
$\xi(\mathbf{r})=exp (\mu(\mathbf{r})\beta(\mathbf{r}))$
Having differentiated the equation for energy conservation over the volume, we obtain  
an interesting relation for the chemical potential, i.e., 
\begin{equation}
\frac{1}{2}\frac{\rho(\mathbf{r})}{\beta(\mathbf{r})}(3-
\sqrt{\beta(r)}\varphi(\mathbf{r}))=\frac{\delta E}{\delta V}\frac{\delta V}{\delta N}=\mu(\mathbf{r})\rho(\mathbf{r})
\end{equation}
which yields the chemical potential to be given by
\begin{equation}
\mu(\mathbf{r})\beta(\mathbf{r})=\frac{3}{2}-\frac{1}{2}\sqrt{\beta(r)}\varphi(\mathbf{r}))
\end{equation}
Within the context of the expression for the density and the definition of  thermal de-Broglie 
wavelength and the gravitation length, i.e., 
\begin{equation}
\Lambda^{-1}(\mathbf{r})=\left(\frac{2 m}{\hbar^{2}\beta(\mathbf{r})}
\right),R_{g}(\mathbf{r})=2\pi G m^{2}\beta(r)
\end{equation}
we can rewrite all the equations and the normalization condition in terms of density and 
temperature. Thus we have
\begin{equation}
\triangle \left(\frac{\Lambda^{3}(\mathbf{r})\rho(\mathbf{r})}{\sqrt{\beta(\mathbf{r})}}\right)+
\frac{R_{g}(\mathbf{r})}{\sqrt{\beta(\mathbf{r})}}\rho(\mathbf{r})=0
\end{equation}
and  the chemical potential reduces to 
\begin{equation}
\mu(\mathbf{r})\beta(\mathbf{r})=\frac{3}{2}-\ln (\Lambda^{3}(\mathbf{r})\rho(\mathbf{r})) 
\end{equation}
In this approach we can obtain too the equation of state for self-gravitating system
if use the thermodynamic relation $P=-\frac{1}{\beta}\frac{\delta S}{\delta V}$ in the 
case energy conservation $E$. In our case, in the definition of ``local entropy'' can use  
the relation $\left(\nabla \varphi(\mathbf{r})\right)^{2}= 
\nabla(\varphi(\mathbf{r})\nabla \varphi(\mathbf{r}))-\varphi(\mathbf{r})\triangle \varphi(\mathbf{r}) $
after that provide the integration on all volume. First part of integration can present as 
surface integral where $\varphi(\mathbf{r})=0$ on the integration surface. After that we can 
present the ``local entropy as
\begin{equation}
S=\int
\left[-\frac{1}{8\pi m^{2}G} \varphi(\mathbf{r})\triangle \varphi(\mathbf{r})- \xi(\mathbf{r})\left(\frac{2\pi
m}{\hbar^{2}\beta(\mathbf{r})}\right)^{\frac{3}{2}}\exp\sqrt{\beta(r)}\varphi(\mathbf{r})\right]d\mathbf{r}
\end{equation}
which can rewrite, using the definition of density of particle, in the the form 
\begin{equation}
S=\int \left[-\rho(\mathbf{r}) \ln (\Lambda^{3}(\mathbf{r})\rho(\mathbf{r}) - \rho(\mathbf{r})\right]d\mathbf{r}
\end{equation}
The local equation of state can present as
\begin{equation}
P(\mathbf{r}) \beta(\mathbf{r}) = \rho(\mathbf{r}) (1-\ln (\Lambda^{3}(\mathbf{r})\rho(\mathbf{r}))=\rho(\mathbf{r}) \left(\mu(\mathbf{r})\beta(\mathbf{r})-\frac{1}{2}\right)
\end{equation}
In the classical case $\Lambda^{3}(\mathbf{r})\rho(\mathbf{r})<< 1$ and $P\beta \bar{=}\rho $ 
but have the multiple which logarithmic dependence from density of particle. Only in the case 
$\Lambda^{3}(\mathbf{r})\rho(\mathbf{r})= 1$ we obtain the equation of state for ideal gas. 
In the case of the ideal gas we obtain usual equation of state, because in this case 
$\varphi(\mathbf{r})=0$ and $P\beta=\rho $ as result absent of interaction. In the case
of ideal gas $\mu(\mathbf{r})\beta(\mathbf{r})=\frac{3}{2}$ and the equation of 
state reproduce the equation of state of the ideal gas. In this case the energy of system 
equal $E=\frac{3}{2}NkT$ that satisfy the previous obtained results. In the  next section 
we find the classical distributions of particles for various inner and external conditions.

\section{Particle and temperature distributions in a self-gravitating system}

\subsection{Homogeneous distribution of particles}

a)First  of all we considere the equilibrium case, with all the parameters being 
independent of space  coordinates. In this case, the energy and total number of 
particles are fixed and, moreover, the temperature and the chemical potential do 
not change in space. Thus the equation for the particle concentration
\begin{equation}
\triangle \left(\frac{\Lambda^{3}(\mathbf{r})\rho(\mathbf{r})}{\sqrt{\beta(\mathbf{r})}}\right)+
\frac{R_{g}(\mathbf{r})}{\sqrt{\beta(\mathbf{r})}}\rho(\mathbf{r})=0
\end{equation}
leads to a simple condition $\sqrt{\beta}\rho(\mathbf{r})=0$  that can be realized 
only for $T\rightarrow \infty$. The particle distribution in a self-gravitating system 
can be homogeneous only  for very high temperatures.

b) Another interesting case is when only particle density depends on the  coordinate  
while the temperature is fixed. In this case the equation for the density take the form
\begin{equation}
\triangle \left(\ln \Lambda^{3}\rho(\mathbf{r})\right)+R_{g}\rho(\mathbf{r})=0
\end{equation}
and can be transformed to 
\begin{equation}
\triangle \left(\ln\rho(\mathbf{r})\right)+R_{g}\rho(\mathbf{r})=0.
\end{equation}
The latter equation has an exact solution $\rho(\mathbf{r})=\frac{2}{R_{g}r^{2}}$ but 
the normalization condition  holds only for the case of a fixed box with size 
$R=\frac{NGm^{2}}{4kT}$, fixed energy $E=NkT$, and the changes of the chemical-potential 
density within the box given by $\mu=kT(\frac{3}{2}-\frac{2 \Lambda^{3}}{4kTR_{g}r^{2}})$. 

As follows from the equation for constant temperature,  the homogeneous distribution of 
particles is unstable. The homogeneous distribution of particle 
$\rho(\mathbf{r})=\rho+\delta \rho(\mathbf{r})$ yields an equation for  density fluctuations, 
i.e.,
\begin{equation}
\triangle \delta\rho(\mathbf{r})+R_{g}\rho \delta \rho(\mathbf{r})=0
\end{equation}
that reproduces the Helmholtz equation.The general solution of the wave equation is the 
unstable radial  distribution $\delta \rho(\mathbf{r})=\frac{exp ikr}{r}$ with the wave 
number $k=\sqrt{2\pi G m^{2} \beta \rho}$ that implies that the wavelength of the instability 
is half as long as the Jeans length. It is the statistical length of the instability of 
particle distribution  in the system.

The concept of Jeans gravitational instability is discussed within the framework of 
non-extensive statistics and its associated kinetic theory \cite{Lim}. A simple analytical 
formula generalizing the Jeans criterion is derived by assuming that the unperturbed 
collisionless gas is kinetically described by the class of power-law velocity distributions. 
It is found that the critical values of the wavelength and mass depend explicitly on the 
non-extensive parameter. The instability condition is weakened  as the system becomes unstable 
even for wavelengths of the disturbance smaller than the standard Jeans length. 

The recent discoveries of extrasolar giant planets, coupled with refined models of the 
compositions of Jupiter and Saturn, prompt a reexamination of the theories of giant 
planet formation. An alternative to the favored core accretion hypothesis is examined 
in \cite{Boss}, the conclusion is that the gravitational instability in the outer solar 
nebula leads to the formation of giant planets. Three-dimensional hydrodynamic calculations 
predict formation with locally isothermal or adiabatic thermodynamics. The gravitational 
instability appears to be capable of forming giant planets \cite{Fall}. Our results can 
help to explain the data of astrophysical observations in the sense that the different 
length of the instability in a self-gravitation system  is associated with the alternative 
description  of the situation. Thus we can conclude that  particle distributions cannot 
be homogeneous for constant temperatures in the system. Thus we have to find real 
distributions of particles and temperature in the system

\subsection{Inhomogeneous distributions of particles and temperature in a self-gravitating system}

In the general case,  particle distributions  in self-gravitating systems are inhomogeneous. 
Inhomogeneous distribution of particles  gives rise to the long-range gravitational interaction.  Now we consider the non-equilibrium description of a self-gravitating system and take into account  probable spatially inhomogeneous distributions of particles and temperature.  We introduce  a new variable $\psi= \Lambda^{3}(\mathbf{r})\rho(\mathbf{r})$, then  the equation for the density is simplfied, i.e., we have 
\begin{equation}
\triangle \left(\frac{\ln \psi}{\sqrt{\beta(\mathbf{r})}}\right)+
\frac{R_{g}(\mathbf{r})}{\sqrt{\beta(\mathbf{r})}\Lambda^{3}(\mathbf{r})}\psi=0
\end{equation}
The solution of  this equation provides a complet non-equilibrium statistical description 
of a self-gravitating system. General exact solutions of this non-linear equation are unknown. 
In what follows we propose a way to solve this equation. 

a) First  of all we can find a more general solution of  the problem. In the three-dimensional 
case the action of the Laplace operator can be presented in the form
\begin{equation}
\triangle \left(\frac{\ln \psi}{\sqrt{\beta(\mathbf{r})}}\right)=
\frac{1}{\sqrt{\beta}}\left(\frac{d^{2}}{dr^{2}}+\frac{2}{r}\frac{d}{dr}\right)\ln \psi-
\frac{\ln \psi}{\sqrt{\beta^{3}}}\left(\frac{d^{2}\beta}{dr^{2}}+\frac{2}{r}\frac{d \beta}{dr}-\frac{3}{2 \beta}(\frac{d\beta}{dr})^{2}\right)-
\frac{1}{\sqrt{\beta^{3}}}\frac{d \ln \psi}{d r}\frac{d \beta}{d r}
\end{equation}
The solution for the temperature  is given by $\beta=\gamma^{3}r^{n}$ 
and it is not difficult to see that for $n=2$ we obtain only an equation for $\psi$, i.e., 
\begin{equation}
\frac{d^{2}\ln \psi}{dr^{2}}+\frac{a_{m}}{B\gamma r}\psi=0
\end{equation}
that can be rewritten in terms of the new variable $\bar{r}^{2}=r$, i.e., 
\begin{equation}
\frac{d}{d\bar{r}}(\frac{1}{\psi}\frac{d\psi}{d\bar{r}})+\frac{4a_{m}}{B\gamma}\psi=0
\end{equation}
We multiply this equation  by $\frac{1}{\psi}\frac{d\psi}{d\bar{r}}$ and calculate 
the first integral of the equation thus obtained.  It is given by
\begin{equation}
(\frac{1}{\psi}\frac{d\psi}{d\bar{r}})^{2}+\frac{4a_{m}}{B\gamma}\psi=\Delta
\end{equation}
and the exact solution can be written as 
\begin{equation}
\psi=\frac{\Delta}{\frac{8a_{m}}{B\gamma}\sinh^{2}\sqrt{\frac{\Delta r}{4}}}
\end{equation}
Using the  latter definition, we  find the exact solution for the inhomogeneous particle 
distribution  to be given by
\begin{equation}
\rho(\mathbf{r})=\frac{\Delta}{8a_{m}\gamma^{2} r^{3}\sinh^{2}\sqrt{\frac{\Delta r}{4}}}
\end{equation}
and thus obtain good assumptions concerning the behavior at long distances  from the 
center of the inhomogeneous particle distribution. This behavior related with result
early obtained in articles \cite{Mich}-\cite{Jaf} where was use Boltzmann equation
for distribution function for spherical isolate stellar system. The distribution of 
particles  is inhomogeneous  for the size $R=\frac{1}{4\Delta}$ and divergent  to center 
$\rho(\mathbf{r})=\frac{1}{2a_{m}\gamma^{2} r^{4}}$. In this case the energy of system
are conserved. However, we do not know the coefficients.  Thus we propose the approach 
given below. If particles are concentrated  at short distances and their concentration 
is very high, the crucial factor is the quantum effect  for which our approach is 
inapplicable. The relation between critical temperature and particle concentration  
in this quantum case is determined  by a natural condition 
\begin{equation}
\Lambda^{3}(\mathbf{r})\rho(\mathbf{r})=\left(\frac{\hbar^{2}\beta_{c}}{2me}\right)^{\frac{3}{2}}\rho_{c}=1
\end{equation}
This relation  along with the formula for the conservation of the number of particles, 
$\frac{4\pi}{2 a_{m}R_{c}}=N$,  determine all the  required parameters, i.e., the critical 
distance 
$R_{c}=\frac{\hbar^{2}}{m a_{m}N^{\frac{1}{3}}}$,  the coefficient 
$\gamma^{2}=\frac{2\pi m e}{\hbar^{2}N^{\frac{2}{3}}}$,  the critical temperature 
$\beta_{c}=\gamma^{2}R^{2}_{c}$, and the concentration $\rho_{c}=\frac{1}{2 a_{m} R^{4}_{c}}$. 
The energy of the system is in this case given by $E=\frac{3}{2}N kT$, i.e., is equal 
to the energy of a free particle! In this part we present the general solution for the 
classical particle distribution  at long distances from the center of an inhomogeneous 
cluster of condensed matter that is subject to the laws of  quantum physics. In all the 
cases, our solution  holds under the condition of classical physics, 
$\Lambda^{3}(\mathbf{r})\rho(\mathbf{r})<< 1$.

b) In this part we describe the system with  
$\Lambda^{3}(\mathbf{r})\rho(\mathbf{r})=\alpha=const<< 1$.
In this case we can determine only the behavior of temperature  that is governed by 
the equation
\begin{equation}
\triangle \left(\frac{1}{\sqrt{\beta(\mathbf{r})}}\right)+
\frac{a_{m}e^{\alpha}}{B \ln \alpha}\frac{1}{\beta}=0
\end{equation}
Similarly to the previous case, we write the solution of this equation in the form 
$\beta=\gamma^{-2}r^{-2n}$ and thus find that it holds for $n=-2$, i.e.,  the temperature 
is changed as $kT=\gamma^{2}r^{-4}$, the concentration is changed as $\rho=Ar^{-6}$ and 
normalization conditions  for the conservation of particle number and energy are satisfied. 
The limiting behavior of the concentration and temperature, similarly to the previous part,
provides a suitable solution of the problem behavior  for this special case. 
Finally, we make an attempt to present an arbitrary solution in the general case of 
space-coordinate-dependences of the concentration and temperature. This equation 
describes any problem associated with inhomogeneous distributions of particles, 
temperature, and concentration in a self-gravitating system. Indeed, though the 
equation cannot be solved in the general case, it provides a possibility to analyze 
many cases of the behavior of a self- gravitating system  under various  external 
condition.  We can consider many realistic  distributions of concentration, temperature, 
and field in a gravitational system but should not use the equation of state, which 
must exist as a condition related with this equation. In a real system, the temperature 
cannot be related to the particle distribution. It is a thermodynamic parameter that 
determines a condition for the behavior of the system and can be found from other 
physical reactions, not only gravitation. In the general case we cannot  obtain the 
general solution of the present equation, but can believe that this equation governs  
the thermodynamics of self-gravitating systems.

Now we consider one of many simple cases. We assume that there exists a polynomial 
relation between density and temperature  that is given by 
$\sqrt{\beta(\mathbf{r})}\rho(\mathbf{r})=A=const$ 
and the density changes in accordance with the non divergence of the number of particles 
that is possible if the density decreases as the fourth power of the distance from center, 
$\rho(r)=Cr^{-4}$. As follows from the normalization condition, $C=NR_{c}$ where $R_{c}$ 
determines the classical limit similarly to the previous case. Such behavior of the density 
is associated with the real behavior of the gravitational field variable. The equation for 
the field variable  yields the field potential $\varphi(\mathbf{r})=\frac{r_{m}A}{r}$ that 
directly reconstructs the behavior of the gravitational field.  As follows from this 
condition, the temperature also decreases as the fourth power of the distance and hence  
$\sqrt{\beta(\mathbf{r})}=\frac{Ar^{4}}{C}$ or $T=\frac{C}{A}r^{-8}$. 
From the equation  for energy conservation, we  obtain a relation between  the constants 
introduced and the energy of the system, i.e.,  
$\frac{C^{2}}{AR^{5}_{0}}-\frac{A C r_{m}}{R^{4}_{0}}=E$. 
In this simple case we can obtain all the required coefficients and the spatial dependence 
of the density and temperature for inhomogeneous particle distributions in a self-gravitating 
system. This solution can describe the behavior of the temperature and concentration in a 
real self-gravitational object provided an analysis of the experimental situation is carried out.

\section{Conclusion}
The self-gravitating systems  are non-equilibrium a priory. Indeed, the  up-to-date 
non-equilibrium statistical description  considers only  probable dilute structures in  
a self-gravitating system but does not describe metastable states and tells nothing about 
the time scales of the kinetic theory. The new approach  in terms of a non-equilibrium 
statistical operator with allowance  for inhomogeneous distributions of particles and 
temperature is proposed. The method employs the saddle-point procedure to find the dominant 
contributions to the partition function and provides a possibility to obtain all the 
thermodynamic parameters of the system. The statistical operator  has no singularities 
for various values of the gravitational field. The approach makes it possible to solve 
the problem of self-gravitating systems of particles  with inhomogeneous distributions 
of particles and temperature. Probable specific features of the behavior of  a self-gravitating 
system are predicted for  various conditions. The equation of state for self-gravitating 
system has been determined. A new length of the statistical instability and parameters 
of the spatially inhomogeneous distribution of particles and temperature are 
found for real gravitational systems. The gravity factor can either promote or retard such 
transformations depending on the system and conditions concerned. For the first time 
a description is given of the formation of spatially inhomogeneous particle distributions 
accompanied by the changes of temperature. The statistical description of the system is 
tailored to treat the gravitating particles with regard for an arbitrary spatially 
inhomogeneous particle distributions. In this approach, the probable behavior of a 
self- gravitating system can be predicted for any external conditions. In this way 
we can solve the complicated problem of statistical description of self-gravitating 
systems. Moreover, the method can also be  applied for the further development of 
physics of self- gravitational and similar systems  that are not far from the equilibrium.

\end{document}